\documentclass[aps,prl,showpacs,amsmath,amssymb,sort&compress,superscriptaddress,reprint,10pt]{revtex4-1}
\usepackage{bm}
\usepackage{graphicx}
\usepackage{xcolor}
\usepackage{times}
\usepackage{bbm}
\usepackage{url}
\usepackage[colorlinks=true,breaklinks=true,allcolors=blue]{hyperref}

\definecolor{jens}{rgb}{0,.8,.5}

\definecolor{michael}{rgb}{1,.4,.3}

\definecolor{mauritz}{rgb}{0,0,1}

\definecolor{salvatore}{rgb}{1,0,1}

\newcommand\NN{{\mathbbm{N}}}

\newcommand\RR{{\mathbbm{R}}}

\newcommand\id{{\mathbbm{1}}}

\newcommand\ee{{\mathrm{e}}}
\newcommand\ii{{\mathrm{i}}}

\DeclareMathOperator{\tr}{tr}
\newcommand\dist{{d}}

\begin{document}

\title{Breakdown of quasi-locality in long-range quantum lattice models}

\author{Jens Eisert} 
\affiliation{\mbox{Dahlem Center for Complex Quantum Systems, Freie Universit\"at Berlin, 14195 Berlin, Germany}} 

\author{Mauritz van den Worm} 
\affiliation{Institute of Theoretical Physics,  University of Stellenbosch, Stellenbosch 7600, South Africa}
\affiliation{National Institute for Theoretical Physics (NITheP), Stellenbosch 7600, South Africa} 

\author{Salvatore R.\ Manmana} 
\affiliation{Institute for Theoretical Physics, Georg-August-Universit\"at G\"ottingen, 37077 G\"ottingen, Germany} 

\author{Michael Kastner} 
\affiliation{Institute of Theoretical Physics,  University of Stellenbosch, Stellenbosch 7600, South Africa}
\affiliation{National Institute for Theoretical Physics (NITheP), Stellenbosch 7600, South Africa}

\date{January 10, 2014}

\begin{abstract}
We study the non-equilibrium dynamics of correlations in quantum lattice models in the presence of long-range interactions decaying asymptotically as a power law. For exponents larger than the lattice dimensionality, a Lieb-Robinson-type bound effectively restricts the spreading of correlations to a causal region, but allows supersonic propagation. We show that this decay is not only sufficient but also necessary. Using tools of quantum  metrology, for any exponents smaller than the lattice dimension, we construct Hamiltonians giving rise to quantum channels with capacities not restricted to a causal region. An analytical analysis of long-range Ising models illustrates the disappearance of the causal region and the creation of correlations becoming distance-independent. Numerical results obtained using matrix product state methods for the XXZ spin chain reveal the presence of a sound cone for large exponents, and supersonic propagation for small ones. In all models we analyzed the fast spreading of correlations follows a power law, but not the exponential increase of the long-range Lieb-Robinson bound.
\end{abstract}


\maketitle 

Owing to the ground breaking progress in the control of ultra-cold atoms and ions, as well as by the creation of ultra-cold polar molecules, a number of proposals and experimental realizations of quantum computers and quantum simulators for strongly correlated, and in particular spin, 
systems have been reported in recent years 
\cite{Struck_etal11,*Islam_etal11,*Islam_etal13,*Yan_etal13,Trotzky_etal12,Britton_etal12}.
In several theoretical proposals, the constituents effectively interact via long-range forces, be it in the form of dipolar interactions 
\cite{DeMille02,*Micheli_etal06,*Lahaye_etal09,*Baranov_etal12,*GorshkovJILA,*Hazzard_etal13,*LongRangeDipolar},
or phonon-mediated effective interactions in ion crystals 
\cite{PorrasCirac04,*Korenblit_etal12}.  
This is different from traditional condensed matter systems where the long-range character of the underlying Coulomb interactions is suppressed by screening so that only on-site, nearest or next-to-nearest neighbor interactions need to be taken into account. Due to their relevance in the context of quantum simulation of many-body systems by ultra-cold gases and trapped ions, long-range interactions have recently moved into the focus of research 
\cite{Kastner10,*KastnerJSTAT10,Kastner11,*Kastner12,GorshkovJILA,*Maik_etal12,*KoffelLewensteinTagliacozzo12,*Wessel,*GongDuan,*Juenemann_etal}.

Of particular interest are those instances where the presence of long-range interactions leads to phenomena that differ drastically from their short-range counterparts. Prime examples are the breakdown of the Mermin-Wagner theorem \cite{MerminWagner66}, or non-equivalent equilibrium statistical ensembles and negative response functions 
\cite{LynWood68,*Thirring70,Kastner10,*KastnerJSTAT10}, or a non-ballistic spread of correlations \cite{HaukeTagliacozzo13} and entanglement \cite{Schachenmayer_etal13}. Out-of-equilibrium, the presence of long-range interactions has been linked to the occurrence of time scales that diverge with system size, implying that equilibrium properties are then effectively unobservable \cite{Kastner11,*Kastner12}. 

In this work we discuss a particularly striking feature of non-relativistic quantum lattice systems with long-range interactions: not only the absence of a group velocity, but the complete disappearance of a causal region. In a non-relativistic theory it may be not much of a surprise that, in the absence of a finite maximum speed, a local change in some part of a spatially extended system can take immediate effect in distant parts. However, at any given time $t$ after the initial perturbation, this effect turns out to be extremely small in the exterior of an effective sound cone, at least for lattice systems with finite-range interactions and finite-dimensional constituents. This effect goes under the name of {\em quasi-locality}\/ and is the content of the Lieb-Robinson theorem, affirming that observables evolving in the Heisenberg picture satisfy 
\begin{multline}\label{e:LRSR}
	\left\lVert\left[O_A(t),O_B(0)\right]\right\rVert\\
	\leq  C \left\lVert O_A\right\rVert \left\lVert O_B\right\rVert 
	\min(|A|, |B|) \ee^{(v\lvert t\rvert- \dist(A,B))/\xi}
\end{multline}
for spin systems on arbitrary lattices (regular graphs) interacting through finite-range or exponentially decaying potentials 
\cite{LiebRobinson72,*NachtergaeleSims10,*KlieschGogolinEisert14,HastingsKoma06,*NachtergaeleOgataSims06}, 
for $C,v,\xi>0$.
Here, $A,B \subset\Lambda$ are non-overlapping regions of the lattice $\Lambda$, and $O_A(0)$ and $O_B(0)$ are 
supported only on the subspaces of the Hilbert space corresponding to $A$ and $B$, respectively. $\dist(A,B)$ denotes the graph-theoretic distance between $A$ and $B$ \footnote{This is the number of edges along the shortest path connecting the two non-overlapping regions.}.
The physical relevance of \eqref{e:LRSR} lies in the fact that it implies an upper bound on the speed at which information, equal-time correlations \cite{BravyiHastingsVerstraete06}, or entanglement \cite{EisertOsborne06} propagate. All these quantities are negligibly small outside the effective causal cone that is determined by those values of $t$ and $\dist(A,B)$ for which the exponential in \eqref{e:LRSR} is larger than some $\epsilon>0$, which happens for $v\lvert t\rvert>\dist(A,B)+{\xi}\ln\epsilon$. The time scale of information propagation also sets an important time scale in equilibration and thermalization in quantum many-body systems out of equilibrium 
\cite{CalabreseCardy06,*CramerEisert08,*Kollath08,*CramerEisertScholl08,*Manmana2009}.

{\em Lieb-Robinson bounds for long-range interactions.---}%
Long-range interactions have a drastic effect on the spreading of correlations. The general form of long-range Hamiltonians is $H_\Lambda = \sum_{X\subset \Lambda} h_X$, where $h_X$ are local Hamiltonian terms of compact support $X$, assumed to satisfy
\begin{equation}\label{condition}
	\sum_{X\ni x,y} \|h_X\| \leq {\lambda_0}{(1+ \dist(x,y) )}^{-\alpha},
\end{equation}
where $\alpha>0$ is the exponent bounding the decay of the interaction strength at large distances, like in the examples \eqref{e:Hchannel}, \eqref{e:H_Ising}, and \eqref{eq:XXZHam} further below. For exponents $\alpha$ larger than the graph theoretical dimension $D$ of the lattice (coinciding with the standard dimension in case of a cubic lattice), Lieb-Robinson-type bounds have been proved
\cite{HastingsKoma06,*NachtergaeleOgataSims06}
\footnote{Eq.~\eqref{e:LRLR} follows from Eq.~(2.2) of \cite{HastingsKoma06} by setting $a = 0$. Bounding the resulting sum over $Y$ by an integral gives the $(\dist(A,B) + 1)^{\alpha-D}$ factor. The sum over $X$ then yields $|X|$ in the bound. Repeating the same calculation with the roles of $X$ and $Y$ interchanged leads to $\min(|X|,|Y|)$ in the upper bound.}
\begin{equation}\label{e:LRLR}
\left\lVert\left[O_A(t),O_B(0)\right]\right\rVert
\leq C \lVert O_A\rVert \lVert O_B\rVert 
\frac{\min(\lvert A\rvert,\lvert B\rvert)  (
e^{v\lvert t\rvert}\!-\!1)
}{(\dist(A,B)+1)^{\alpha-D}}.
\end{equation}
Analogous to the short-range case, this expression limits the way information can propagate in a quantum lattice system, and we obtain an effective {\it causal region} by determining those values of $t$ and $\delta$ for which the fraction on the right-hand side of \eqref{e:LRLR} is larger than $\epsilon>0$. This condition gives rise to the inequality
\begin{equation}\label{e:causalregion}
	v\lvert t\rvert>\ln\biggl[1+\frac{{\epsilon}(1+\dist(A,B))^{\alpha-D}}{\min(|A|,|B|)}\biggr],
\end{equation}
resulting in an effective causal region that is growing logarithmically for large distances. In contrast to the short-range case, spreading of correlations (or entanglement, or information) is not confined to the interior of a linear sound cone, an effect we refer to as ``supersonic'' propagation. The above discussion of a supersonic causal region is based on the bound \eqref{e:LRLR}, proved only for $\alpha>D$.

In this work we show that an exponent $\alpha$ of this magnitude is not only sufficient, but also necessary, for having a causal region. For long-range models on arbitrary lattices, with interactions decaying asymptotically like $1/\delta^\alpha$ as a function of the distance $\delta$, we construct, for every $\alpha\leq D$, models and entangled initial states such that no causal region exists. We do so by combining methods of many-body theory with those of quantum information and quantum metrology. We also give evidence that, for product initial conditions, a causal region can be identified for $\alpha>D/2$. We provide an analytical analysis of the causal region of the long-range Ising model, as it is relevant for ion-trap systems. Using matrix product state-based methods we further corroborate these findings by an analysis of correlations in long-ranged XXZ chains, which, in contrast to the Ising model, possess strong quantum fluctuations and hence support the validity of our findings for a wider range of situations.

{\em Nonequilibrium systems as quantum channels.---}%
In\-form\-ation propagation can most naturally be captured by capacities of quantum channels, reflecting time evolution. The most appropriate concept here is the classical information capacity. At $t=0$, the system is prepared in the state $\rho$. The coding amounts to either implementing a quantum channel
\begin{equation}
	\rho\mapsto T_t(\rho):= \tr_{{\Lambda}\backslash B} 
	(e^{-\ii tH} U_A\rho U_A^\dagger e^{\ii tH}), 
\end{equation}
for $t\geq 0$, where $U_A$ is non-trivially supported on $A$ only, or
\begin{equation}
	\rho\mapsto N_t(\rho):= \tr_{{\Lambda}\backslash B} (e^{-\ii tH}\rho e^{\ii tH}),
\end{equation}
merely reflecting free time evolution. Then one performs a measurement associated with a positive-operator valued measure $\pi_B$ supported on $B$ only, with $0\leq \pi_B\leq \id$. The classical information capacity $C_t$ can, in the setting considered here, be bounded from below by the probability of detecting a signal at time $t>0$, so 
\begin{equation}
	C_t \geq p_t:= \left\lvert \tr\left(T_t(\rho) \pi_B\right) - \tr\left(N_t(\rho) \pi_B\right)\right\rvert.
\end{equation}
This channel capacity $C_t$ captures the rate at which classical bits of information can be reliably transmitted from $A$ to $B$, by either performing a local unitary $U_A$ at $t=0$ or not, and detecting the signal in $B$ at a later time.

{\em Lower bounds on information propagation from product states.---}%
We consider a family of lattices $\Lambda$ of arbitrary dimension $D$ with open boundary conditions. For our purposes, we are free to consider any Hamiltonian that is compatible with Eq.\ (\ref{condition}). Here we choose
\begin{equation}\label{e:Hchannel}
	H_\Lambda 
	= \frac{1}{2}(\id - \sigma^z_o ) 
	\sum_{j\in B} \frac{1}{(1+ \dist(o,j))^\alpha}
	(\id - \sigma^z_j ),
\end{equation}
resembling an Ising interaction, where $o\in \Lambda$. Here, $\sigma^z$ is the $z$ component of the Pauli spin operator supported on site $j$. We assume $A=\{o\}$, so that $|A|=1$ and take $B:= \{ j\in \Lambda:  \dist(o,j)\geq  \delta\}$  for some $\delta\in \NN$. As initial state, we choose $\rho=|0\rangle\langle 0|^{\otimes|\Lambda\setminus B|}\otimes |+\rangle\langle +|^{\otimes |B|}$ with $|+\rangle=(|0\rangle + |1\rangle)/\sqrt{2}$, and the first tensor factor corresponds to the complement $\Lambda\backslash B$ of $B$ on the lattice. We choose $\pi_B = |+\rangle\langle +|^{\otimes |B|}$ and $U_A=|1\rangle\langle 0| $. In these terms,
\begin{equation}
p_t= 1- \frac{1}{2^{|B|}}\prod_{j\in B}\biggl[1+\cos\Bigl(\frac{2t}{ (1+\dist(o,j) )^\alpha   }\Bigr)\biggr].
\end{equation}	
For times $t$ such that $2t\leq (1+\delta)^\alpha$, we obtain (for details, see Supplemental Material)
\begin{equation}\label{e:pt}
	p_t\geq 1-  \exp\biggl[-\frac{ 4 t^2}{5}\sum_{j\in B } (1+ \dist(o,j))^{-2\alpha}\biggr].
\end{equation}
We denote with $O_{\Lambda,l}$ the number of sites $j$ in the lattice $\Lambda$ for which $\dist(o,j)=l$. By definition $O_{\Lambda,l}=\Theta(l^{D-1})$, which means it is up to constants linearly upper and lower bounded by $l^{D-1}$ \footnote{For example, the number of sites on the surface of a hypercube of side length $2l$ in a cubic lattice of dimension $D$ is $2 d (2l)^{D-1}$.}. 
Expressing the sum in \eqref{e:pt} in terms of this quantity, one finds
\begin{eqnarray}
	\sum_{j\in B} 
	\left(
	1+ \dist(o,j)
	\right)^{-2\alpha} = \sum_{l=\delta}^L \left(
	1+ l
	\right)^{-2\alpha} O_{\Lambda,l}.
\end{eqnarray}
The right hand side converges exactly if 
             $\lim_{L\rightarrow\infty}  \sum_{l=\delta}^{{L}} (1+l)^{{D-1}-2\alpha}<\infty$,
which is when {$\alpha>D/2$}. Hence, for $\alpha<D/2$ and for the product initial states considered, for any constant $c\in [0,1]$ and sufficiently small times $t>0$, for a sufficiently large $|B|$ (and lattice), one can detect a signal $C_t>c$. Signal propagation is therefore not restricted to any causal region, as reliable information can be transmitted arbitrarily fast, beyond any finite speed of information propagation. Counterexamples of cone-like propagation are also known from lattice systems with nearest-neighbor interactions and infinite-dimensional constituents \cite{EisertGross09}.

{\em Information spread for multi-particle entangled states.---}%
The bound can yet be beaten by resorting to ideas of metrology and methods of multi-particle entanglement: We will show that, precisely for $D\geq \alpha$, the causal region generally disappears. The steps used here are borrowed from
quantum phase estimation. We start with the initial state $\rho=|0\rangle\langle 0|^{|\Lambda|-|B|}\otimes |\psi\rangle\langle\psi|$, with 
\begin{equation}
	|\psi\rangle = (|0,\dots, 0\rangle + |1,\dots ,1\rangle)/\sqrt{2}.
\end{equation}
That is, the subset $B\subset\Lambda$ of the lattice is prepared in a multi-partite entangled Greenberger-Horne-Zeilinger state---of the kind as it is used in error-free phase estimation and metrology \cite{Frequency}. When the system is prepared in this state, but otherwise in the same situation as considered before, one finds
\begin{equation}
	p_t = 1-  \frac{1}{2}
	\biggl[1+ \cos\biggl( t \sum_{j\in B}
	(1+ \dist(o,j))^{-\alpha} 
	\biggr)
	\biggr].
\end{equation}
Again expressed in terms of the number of lattice sites at a given distance,
\begin{equation}
	p_t = 1-  \frac{1}{2}
	\biggl[1+ \cos\biggl(t \sum_{l=\delta}^L
	(1+ l)^{-\alpha} O_{\Lambda,l}
	\biggr)
	\biggr].
\end{equation}
Since $1- (1+\cos(x))/2 = x^2/4 + O(x^3)$, we need to further investigate $f(\delta):=\lim_{L\rightarrow\infty}\sum_{l=\delta}^L (1+ l)^{-\alpha} O_{\Lambda,l}$. Exploiting the asymptotic behavior of the Hurwitz zeta function (for details, see the Supplemental Material), one finds
\begin{equation}\label{e:f}
	f(\delta)^{2}= \Theta (\delta^{2(D-\alpha)}).
\end{equation}
This expression defines the asymptotic shape of the causal region for this model. Applied to $\alpha>D$, Eq.\ \eqref{e:f} gives rise to a bent causal region and allows for faster than linear propagation of information, but slower than in principle permitted by Eq.\ \eqref{e:causalregion}. Remarkably, the use of entangled input states alters the threshold of the exponent $\alpha$ from $D/2$ to the optimal value of $D$: On arbitrary lattices, and precisely for $\alpha> D$, the Lieb-Robinson bound \eqref{e:LRLR} sets in and defines a causal region. For $\alpha\leq D$, in contrast, models can be identified that exhibit no causal region at all. This observation does not mean that the Hamiltonian is unphysical---it only implies that, for algebraically decaying interactions, the familiar picture of the existence of causal regions can be drastically altered. These results also generalize and complete findings of Ref.\ \cite{HaukeTagliacozzo13} (where, for $D=1$ and $\alpha<1$, also instances of instantaneous transmission of information have been observed) and complement recent insights into the growth of the mutual information and bi-partite entanglement following quenches \cite{Schachenmayer_etal13}.

{\em Long-range quantum Ising model.---}%
Having established that suitably constructed Hamiltonians can indeed give rise to supersonic propagation and a power law-shaped causal region, we show that such behavior also occurs in the long-range versions of two popular spin models. Moreover, we investigate in this context how correlations spread for exponents $\alpha\leq D$ where the bound \eqref{e:LRLR} is not valid.

\begin{figure}\centering
\includegraphics[width=0.32\linewidth]{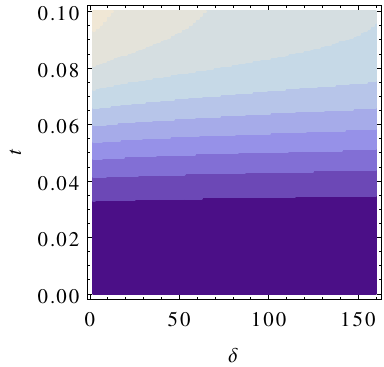}
\includegraphics[width=0.32\linewidth]{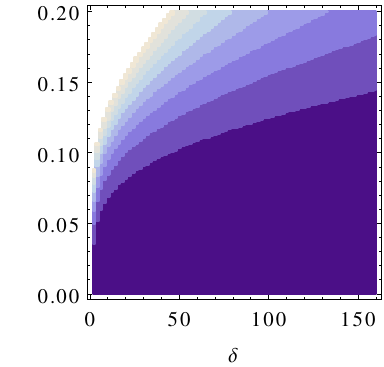}
\includegraphics[width=0.32\linewidth]{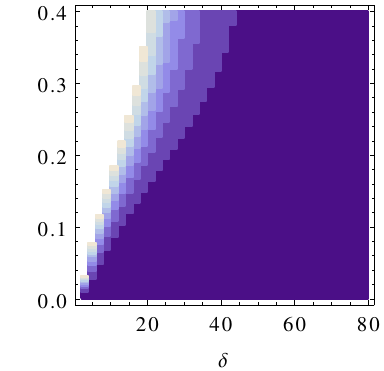}
\caption{\label{f:IsingDensity}%
(Color online) Density contour plots of the connected correlator $\left\langle\sigma_o^x\sigma_\delta^x\right\rangle_\text{c}$ in the $(\delta,t)$-plane for long-range Ising chains with $\lvert\Lambda\rvert=1001$ sites and three different values of $\alpha$. Dark colors indicate small values, and initial correlations at $t=0$ are vanishing.
}%
\end{figure}

The first model we consider is a long-range interacting Ising model on a lattice $\Lambda$ of dimension $D$,
\begin{equation}\label{e:H_Ising}
H_\Lambda=-\sum_{i\neq j}J_{i,j}\sigma_i^z\sigma_j^z,
\end{equation}
where $J_{i,j}\in \RR$ denotes the coupling strength in terms of the distance between the lattice sites $i$ and $j$. Despite the fact that all terms in the Hamiltonian \eqref{e:H_Ising} commute, rich dynamics and genuine quantum effects have been reported to occur 
\cite{Emch66,*Radin70,*Kastner11,*Kastner12,*BachelardKastner13,*Hazzard_etal13,*FossFeigPRA2013,*FossFeig_etal13,vdWorm_etal13}. 
In Ref.\ \cite{vdWorm_etal13}, analytic formul{\ae} for the time evolution of spin--spin correlation functions have been derived for arbitrary lattices, boundary conditions, and couplings, and for a large class of product initial states (Eqs.\ (5a)--(5d) in \cite{vdWorm_etal13}).
We evaluated these formul{\ae} for a chain of spins ($D=1$) with interactions $J_{i,j}=J \dist(i,j)^{-\alpha}$. The spreading of correlations is investigated by plotting, in Fig.\ \ref{f:IsingDensity}, the density contours of the connected correlator $\left\langle\sigma_o^x\sigma_\delta^x\right\rangle_\text{c}=\left\langle\sigma_o^x\sigma_\delta^x\right\rangle-\left\langle\sigma_o^x\right\rangle\left\langle\sigma_\delta^x\right\rangle$ in the $(\delta,t)$-plane.

For $\alpha=1/4$ (left panel of Fig.\ \ref{f:IsingDensity}) correlations increase in an essentially distance-independent way. A finite-size scaling analysis (see Supplemental Material) confirms that the propagation front indeed becomes flat ($\delta$-independent) for $0\leq\alpha<D/2$, and hence no effective causal region is present. This is consistent with the fact that, for these values of $\alpha$, the Hamiltonian \eqref{e:H_Ising} does not meet the conditions required for the proof of the long-range Lieb-Robinson bound \eqref{e:LRLR}. For $\alpha=3/4$ (central panel of Fig.\ \ref{f:IsingDensity}) the spreading of correlations shows a distance-dependence that is consistent with a power law-shaped causal region; plots for other $D/2<\alpha<D$ are similar. These findings nicely match our results on the channel capacity of \eqref{e:Hchannel}, namely that, for product initial states, an effective causal region is present already for $\alpha>D/2$, and not only for $\alpha>D$ as in the more general case of entangled initial states. For $\alpha=3/2$ (right panel of Fig.\ \ref{f:IsingDensity}) correlations initially seem to spread linearly, but not further than a few tens of lattice sites; plots for other $\alpha>D$ are similar. The breakdown of the initial linear spread in Fig.\ \ref{f:IsingDensity} (right) is presumably a peculiarity of the long-range Ising model and may be explained by the fact that quasi-particles in a spin-wave approximation are dispersionless for the Hamiltonian \eqref{e:H_Ising} \cite{HaukeTagliacozzo13}.

\begin{figure}\centering
\includegraphics[height=0.30\linewidth]{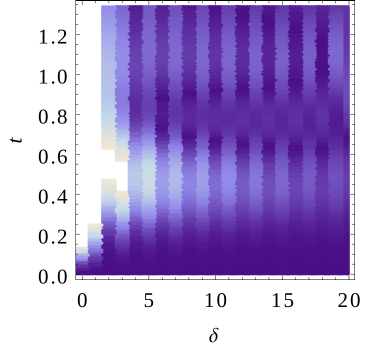}
\hspace{-3.5mm}
\includegraphics[height=0.304\linewidth]{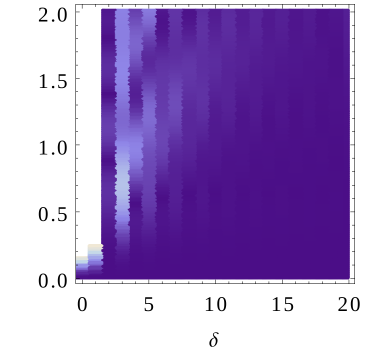}
\hspace{-3.5mm}
\includegraphics[height=0.304\linewidth]{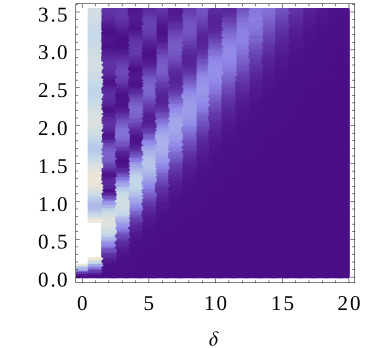}
\includegraphics[height=0.308\linewidth]{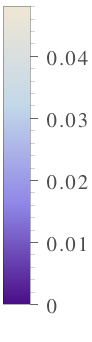}
\includegraphics[height=0.30\linewidth]{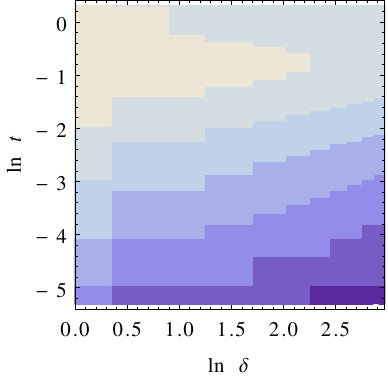}
\hspace{-3.3mm}
\includegraphics[height=0.304\linewidth]{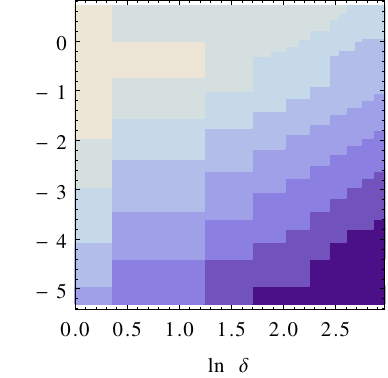}
\hspace{-3.3mm}
\includegraphics[height=0.304\linewidth]{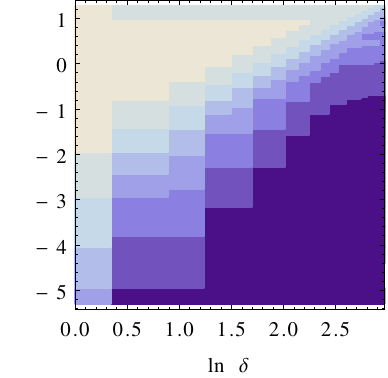}
\hspace{0.5mm}
\includegraphics[height=0.308\linewidth]{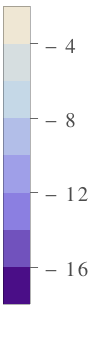}
\caption{\label{f:XXZ}%
(Color online) Top row: Density plots of the correlator $\left\langle \sigma^z_o \sigma^z_\delta\right\rangle_\text{c}$ in the $(\delta,t)$-plane. The results are for long-range XXZ chains with $\lvert\Lambda\rvert=40$ sites and exponents $\alpha=3/4$, $3/2$, and 3 (from left to right).
The left and center plots reveal supersonic spreading of correlations, not bounded by any linear cone, whereas such a cone appears in the right plot for $\alpha=3$. Bottom row: As above, but showing contour plots of $\ln\left\langle \sigma^z_o \sigma^z_\delta\right\rangle_\text{c}$ in the $(\ln\delta,\ln t)$-plane. For better visualization, odd/even effects caused by the staggered initial state have been eliminated (see Supplemental Material for details). All plots in the bottom row are consistent with a power law-shaped causal region for larger distances $\delta$.
}%
\end{figure}

{\em Numerical results on the long-range XXZ spin chain.---}%
To investigate how the observations of the preceding section are affected by the presence of dispersion, we augment the Ising Hamiltonian \eqref{e:H_Ising} with a non-commuting term. We do this by adding interactions in the transverse direction, leading to the long-range $S=1/2$ XXZ chain
\begin{equation}\label{eq:XXZHam}
H^\text{XXZ} = \sum_{i > j} \frac{1}{\dist(i,j)^{\alpha}} \left[\frac{J_{\perp}}{2} \left( \sigma_i^+ \sigma_j^- + \sigma_i^- \sigma_j^+ \right) + J_z \, \sigma_i^z \sigma_j^z \right]. 
\end{equation}
For short-ranged interactions, this is a standard model for the investigation of quantum magnetism. Here, we choose $J_{\perp} = 2$ and $J_z = 1$, so that we are dealing with strong quantum fluctuations. In Fig.\ \ref{f:XXZ} we show numerical results for the time evolution under the Hamiltonian \eqref{eq:XXZHam} with $\alpha = 3/4$, $3/2$, and $3$, starting from the staggered initial state vector $|\psi_0 \rangle = |1,0,1,0,\dots, 1,0  \rangle$ (reminding of the situation of Ref.\ \cite{Trotzky_etal12}). For details of the time-dependent density matrix renormalization group calculations, see the supplemental material. The plots show results for the equal-time connected correlation functions $\left\langle \sigma^z_o \sigma^z_\delta\right\rangle_\text{c} = \langle \sigma_o^z \sigma_\delta^z \rangle - \langle \sigma_o^z \rangle \langle \sigma_\delta^z\rangle$, and similar plots for the correlator $\langle \sigma_o^+ \sigma_\delta^-\rangle$ are provided in the supplemental material.

For $\alpha=3/4$ and $3/2$ (left and middle column in Fig.\ \ref{f:XXZ}) we observe, similarly to the Ising case, a faster-than-linear creation of correlations that appears consistent with a power law-shaped causal region, and even the corresponding timescale is comparable for both models. For $\alpha=3$ (right column in Fig.\ \ref{f:XXZ}) the spreading of correlations is predominantly linear in the $(\delta,t)$-plane. Zooming in (bottom panels of Fig.\ \ref{f:XXZ}) reveals that in this case information appears to leak into the region outside of the cone. This is interesting and hints at the possible coexistence of supersonic propagation and a sound-cone. However, the signal is very small and could be affected by numerical errors (see the Supplemental Material),  so that we leave this issue to future investigations. Numerical results for $\alpha=1/4$ (not shown here) look qualitatively similar to those for $\alpha=3/4$, although the propagation is faster. The accessible chain lengths are too short to distinguish distance-dependent propagation (expected for $\alpha>D/2$) from distance-independent propagation (expected for $\alpha<D/2$), and a finite-size scaling analysis of several longer chains would be required.

{\em Conclusions.---}In this work we have addressed how long-range interactions in quantum lattice models affect the spreading of correlations, entanglement, and quantum information. For interactions decaying asymptotically as $1/\delta^\alpha$ with exponents $\alpha>D$, a causal region can be identified. This causal region is not necessarily cone-shaped, but propagation can be faster than linear. For all models studied we found a power-law shaped causal region, i.e., effects exceed some small $\epsilon>0$ for $v|t|>\epsilon\delta^q$ with $v,q>0$. In contrast, for any $\alpha\leq D$, we constructed models and entangled initial states such that the spreading of correlations is not restricted to any causal region. In fact, in a physical realization of such a situation, one would eventually be constrained by relativistic causality, but this happens only on time and length scales very different from those usually probed in condensed matter or cold atomic realizations of quantum lattice models. Studying the same models, but starting from product initial states, we find a causal region for all $\alpha>D/2$. The findings are corroborated by analytical results on long-range Ising models and numerical results on long-range XXZ models from density matrix renormalization group calculations, both for product initial states. Based on the evidence given, one may conjecture whether generally, for any $\alpha>D/2$ and product initial states, a causal region exists.

The findings presented here suggest that the non-equi\-lib\-ri\-um dynamics of quantum many-body systems with long-range interactions possesses exciting and peculiar features which we expect to be observable in ion-trap quantum simulators \cite{Britton_etal12}. Yet some interesting questions remain open: E.g., while the power law-shaped causal regions we found are not inconsistent with the long-range Lieb-Robinson bound \eqref{e:LRLR}, the actual propagation is much slower than what is permitted by the bound. It would be desirable to either find interacting models with finite-dimensional constituents that give rise to a causal region which widens exponentially, or to prove that such models do not exist. It is our hope that the present work will stimulate further studies on this topic. 

M.\,K.\ acknowledges support by the Incentive Funding for Rated Researchers program of the National Research Foundation of South Africa, J.\,E.\ by the EU (Q-Essence, SIQS, RAQUEL), and the ERC (TAQ). The authors have profited from useful discussions with F.\ Essler, A.\ Gorshkov, A.\ Mitra, A.\ M.\ Rey, and especially from inspiring conversations with M.\ Friesdorf and F.\ Verstraete.

\newpage

\begin{center}
{\bf Supplemental Material}  
\end{center}
\vspace{-3mm}

\appendix
\numberwithin{equation}{section}
\numberwithin{figure}{section}

\section{A. Details of the lower bounds on information propagation}
\setcounter{section}{1}
\setcounter{equation}{0}
\setcounter{figure}{0}
In this section we present details of the argument leading to lower bounds on information propagation. In Eq.\ (9), capturing the situation of using product initial states, the following argument is applied: For times $t$ such that 
 \begin{equation}
 	2t\leq (1+\delta)^\alpha, 
\end{equation}
we can make use of the fact that 
\begin{equation}
	\cos(x)\leq 1-2 x^2/5 
\end{equation}
for all $x\in [0,1]$, to get
\begin{equation}
	p_t\geq 1-  \prod_{j\in B} \left[ 1-\frac{4 t^2}{5(1+ \dist(o,j))^{2\alpha}}\right].
\end{equation}
Using that $\ln(1-x)\leq -x$ for all $x\geq0$, we find
\begin{equation}
	p_t\geq 1-  \exp\biggl[-\frac{ 4 t^2}{5}\sum_{j\in B } (1+ \dist(o,j))^{-2\alpha}\biggr],
\end{equation}
which is the expression given.

\setcounter{section}{2}
\begin{figure}[b]\centering
\includegraphics[width=0.48\linewidth]{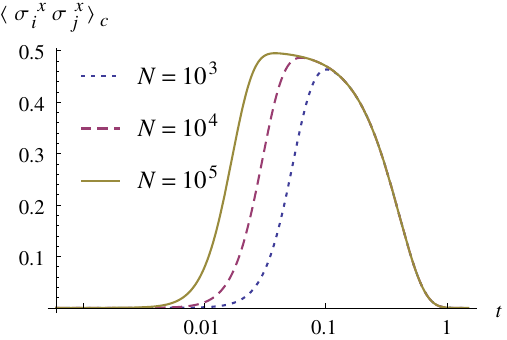}
\includegraphics[width=0.48\linewidth]{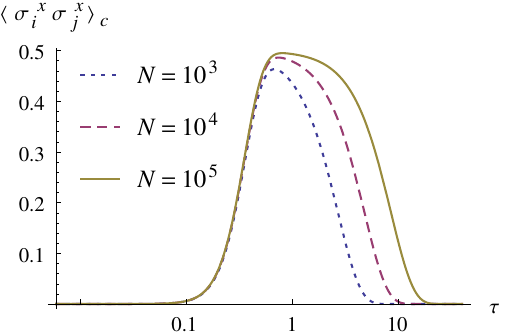}
\caption{\label{f:NDependence}%
(Color online) Time-dependence of the connected correlator $\left\langle\sigma_o^x\sigma_\delta^x\right\rangle_\text{c}$ of the long-range Ising model with $\alpha=1/4$ and $\delta=20$ for system sizes $N=10^3$, $10^4$, and $10^5$. Left panel: As a function of time $t$, the time scale on which correlations build up is strongly $N$-dependent and even vanishes with increasing $N$. The decay of correlations takes place on a time scale that is independent of $N$. Right panel: When plotted as a function of rescaled time $\tau=tN^{1/2-\alpha}$, the time scale on which correlations build up becomes independent of the system size $N$.
}%
\end{figure}
\setcounter{section}{1}

One can similarly proceed in order to arrive at the conclusion of Eq.\ (15) in case of entangled initial states. For a suitable function $g:\RR^+\times \RR^+\rightarrow \RR^+$, we have
\begin{equation}
	f(\delta) \geq   g(\alpha,D)\lim_{L\rightarrow\infty}\sum_{l=\delta}^L
	  l^{D-1-\alpha} 	
=g(\alpha,D) \zeta(\alpha-D+1,\delta)	
\end{equation}	
in terms of the Hurwitz zeta function $\zeta$. Using the asymptotic behavior of the Hurwitz zeta function in the second argument, one finds
\begin{equation}
	\zeta(x,\delta)=\Theta (\delta^{-x+1})
\end{equation}
for any $x\in \RR$, which proves Eq.\ (15).

\section{B. Finite-size scaling analysis of the propagation of correlations in the long-range Ising model with \texorpdfstring{$0\leq\alpha<D/2$}{0<=alpha<D/2}}
\setcounter{section}{2}
\setcounter{equation}{0}
\setcounter{figure}{0}

As can be seen in Fig.\ 1 (left panel) of the main paper, correlations in the long-range Ising chain with $\alpha=1/4$ spread in a broad, more or less distance-independent front, and qualitatively similar results are found for all $0\leq\alpha<D/2$. Closer scrutiny, however, reveals a minuscule decrease of correlations with increasing distance $\delta$. In this section we show that, for $0\leq\alpha<D/2$, this $\delta$-dependence is a finite-size effect that disappears in the large-system limit.

\setcounter{figure}{1}
\begin{figure}\centering
\includegraphics[width=0.48\linewidth]{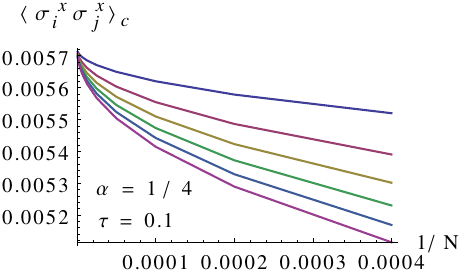}
\includegraphics[width=0.48\linewidth]{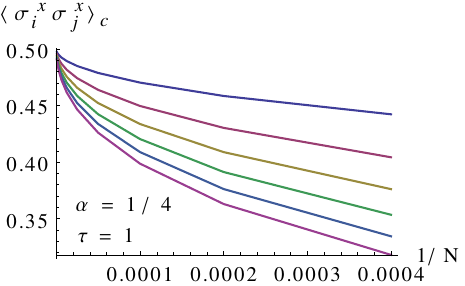}
\includegraphics[width=0.48\linewidth]{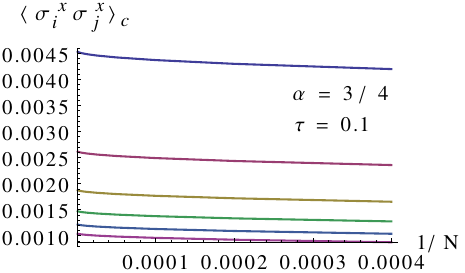}
\includegraphics[width=0.48\linewidth]{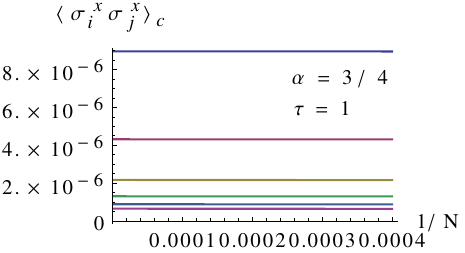}
\caption{\label{f:CorrNDependence}%
(Color online) The connected correlator of the long-range Ising model for fixed values of the rescaled time $\tau=0.1$ (left panels), and $\tau=1$ (right panels). In each plot, $\left\langle\sigma_o^x\sigma_\delta^x\right\rangle_\text{c}$ is shown for various distances $\delta=20$, 40, 60, 80, 100, and 120 (from top to bottom line in each graph) and plotted {\em vs.}\ the inverse system size $1/N$. Top panels: For $\alpha=1/4$ and in the limit $1/N\to0$, the connected correlators become independent of the distance $\delta$, implying a flat (distance-independent) propagation front. Bottom panels: For $\alpha=3/4$, the distance-dependence does not vanish in the large-system limit.
}%
\end{figure}

When plotting the connected correlator $\left\langle\sigma_o^x\sigma_\delta^x\right\rangle_\text{c}$ as a function of time, we observe that, for exponents $0\leq\alpha<D/2$, the time scale on which correlations build up depends strongly on the system size $N:=|\Lambda|$ and vanishes with increasing $N$ (Fig.\ \ref{f:NDependence}). From an upper bound on correlation functions derived and discussed in Ref.\ \cite{vdWorm_etal13}, this time scale is known to scale like $N^{\alpha-1/2}$ asymptotically for large $N$. Plotting $\left\langle\sigma_o^x\sigma_\delta^x\right\rangle_\text{c}$ {\em vs.}\ the rescaled time $\tau=tN^{1/2-\alpha}$ absorbs the $N$-dependence of the time scale. Only in rescaled time $\tau$ it therefore makes sense to investigate the large-system asymptotic behavior of the shape of the propagation front.

In Fig.\ \ref{f:CorrNDependence} we plot, for some fixed instance of $\tau$ and various values of $\delta$, the connected correlator $\left\langle\sigma_o^x\sigma_\delta^x\right\rangle_\text{c}$ as a function of the inverse system size $1/N$. In the limit $1/N\to0$, the correlation function converges to a value that is independent of the distance $\delta$, implying a flat (distance-independent) propagation front. Similar plots are obtained and identical conclusions can be drawn for other exponents $0\leq\alpha<D/2$ and other values of the rescaled time $\tau$.

\section{C. Details of the numerical procedure}
\setcounter{section}{3}
\setcounter{equation}{0}
\setcounter{figure}{0}

We use a Krylov variant of the adaptive time-dependent DMRG based on matrix-product states
\cite{Daley2004,White2004,Schmitteckert2004,Feiguin2005,Manmana2005} which allows us to treat interactions of arbitrary range. 
We choose a time-step of $\delta t = 0.0025$ and keep up to 500 density matrix eigenstates, aiming for a discarded weight of $10^{-9}$ or smaller during the time evolution.
We prepare the initial state by applying a staggered external magnetic field on a non-interacting system, which leads to a product state as discussed in the main text, and then perform the time evolution with the long-range XXZ Hamiltonian.  
Due to the stronger increase of entanglement it becomes increasingly difficult to reach longer times the smaller the exponent $\alpha$. 
However, we are able to reach time scales long enough to identify the most prominent features in the time evolution of the correlation functions.  
In order to estimate error bars, we use the discarded weight $\varepsilon$ as a measure for the accuracy of the runs.

As discussed in Ref. \cite{White2007}, for ground state calculations the error is of the order $\sqrt{\varepsilon}$, i.e., in our simulations the accuracy of the observables should be of the order of $\sim 10^{-4}$. 
However, we are treating systems out-of-equilibrium, so it is unclear to which extent the variational argument of Ref.\ \cite{White2007} holds. 
We therefore use as conservative estimate of $10\sqrt{\varepsilon}$ for the absolute error of our observables, i.e., the assumed error bar at the end of the time evolution for $\alpha=3$ and $\alpha=3/2$ is of the order of $10^{-3}$, and for $\alpha = 3/4$ it is of the order of $10^{-2}$. 
Note that, following this point of view, at shorter times the accuracy is much higher. 
Since the signal in the correlation functions is of the order of 0.1, this accuracy is high enough to determine accurately the position of the horizon.

\section{D. Further numerical results on the XXZ model}
\setcounter{equation}{0}

\setcounter{section}{4}
\setcounter{figure}{0}
\begin{figure}\centering
\includegraphics[height=0.295\linewidth]{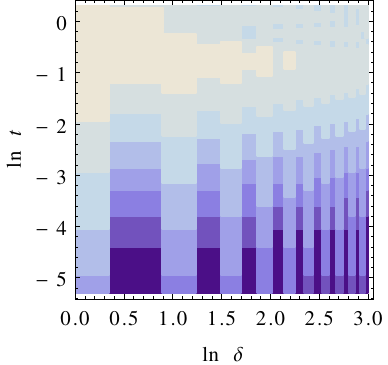}
\hspace{-3.2mm}
\includegraphics[height=0.295\linewidth]{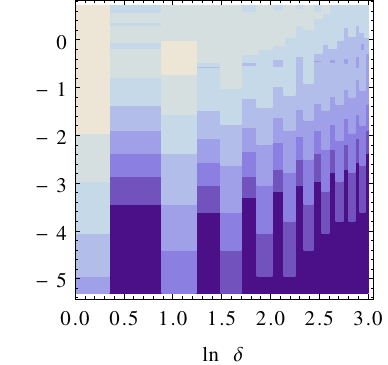}
\hspace{-3.2mm}
\includegraphics[height=0.295\linewidth]{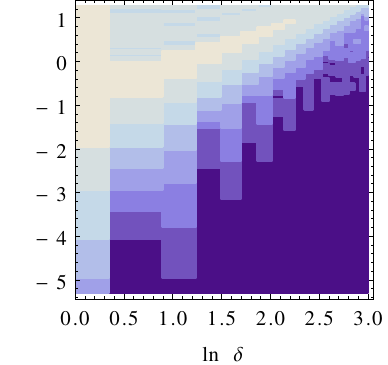}
\includegraphics[height=0.30\linewidth]{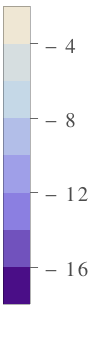}
\caption{\label{f:XXZoriginal}%
(Color online) Density plots of the logarithm of the correlator $\left\langle \sigma^z_o \sigma^z_\delta\right\rangle_\text{c}$ in the $(\ln\delta,\ln t)$-plane. The results are for long-range XXZ chains with $\lvert\Lambda\rvert=40$ sites and exponents $\alpha=3/4$, $3/2$, and 3 (from left to right).
}
\end{figure}

As mentioned in the caption of Fig.\ 2 in the main paper, we have eliminated in the bottom row of that figure the odd/even effects caused by the staggered initial states. The original data for the connected correlation functions $\left\langle \sigma^z_o \sigma^z_\delta\right\rangle_\text{c}$ are shown in Fig.\ \ref{f:XXZoriginal}. From these plots it is difficult to discern the shape of the causal region. For this reason, we have eliminated the staggering by computing, for fixed time $t$, the smallest possible monotonically decreasing upper bound on the numerical data,
\begin{equation}\label{e:max}
\overline{\left\langle \sigma^z_o \sigma^z_\delta\right\rangle}_\text{c}(t)=\max_{i\geq\delta}\left\langle \sigma^z_o \sigma^z_i\right\rangle_\text{c}(t),
\end{equation}
and it is this quantity that is shown in the bottom row of Fig.\ 2 in the main text. 
For the purpose of comparing the numerical data to a spatially monotonically decreasing function, the maximization in \eqref{e:max} is harmless.

\begin{figure}\centering
\includegraphics[height=0.30\linewidth]{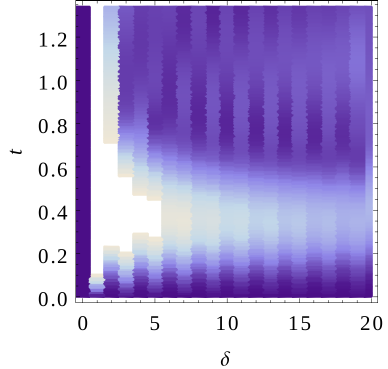}
\hspace{-3.5mm}
\includegraphics[height=0.304\linewidth]{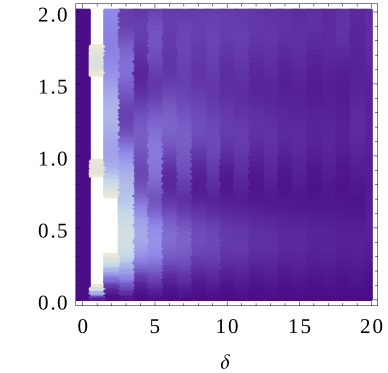}
\hspace{-3.5mm}
\includegraphics[height=0.304\linewidth]{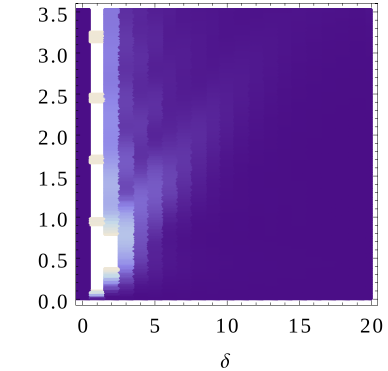}
\includegraphics[height=0.308\linewidth]{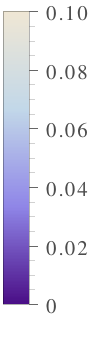}
\includegraphics[height=0.295\linewidth]{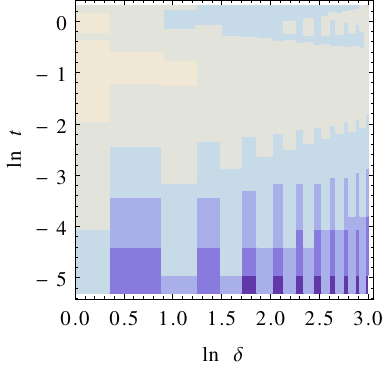}
\hspace{-3.2mm}
\includegraphics[height=0.295\linewidth]{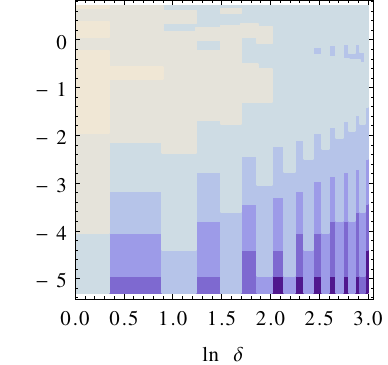}
\hspace{-3.2mm}
\includegraphics[height=0.295\linewidth]{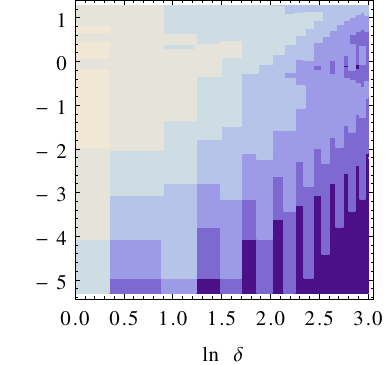}
\includegraphics[height=0.30\linewidth]{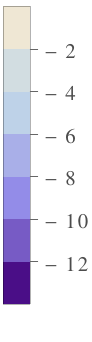}
\includegraphics[height=0.3\linewidth]{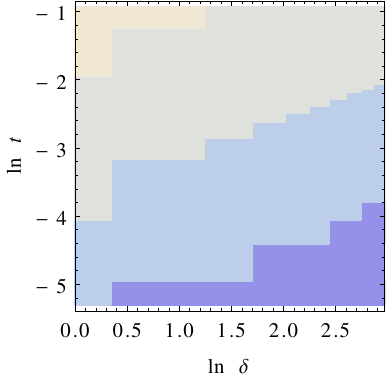}
\hspace{-3.1mm}
\includegraphics[height=0.3\linewidth]{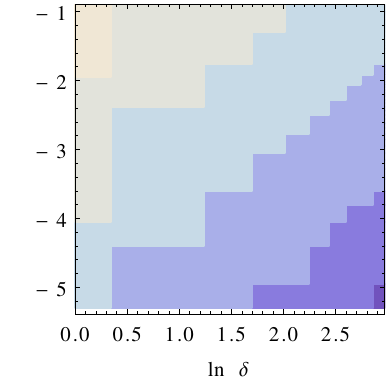}
\hspace{-3.mm}
\includegraphics[height=0.298\linewidth]{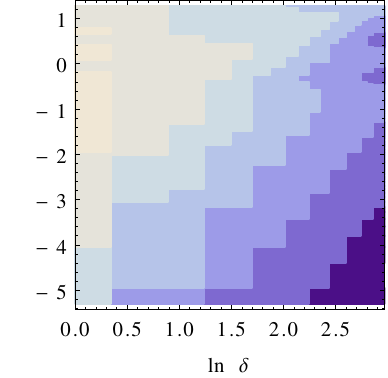}
\hspace{0.3mm}
\includegraphics[height=0.303\linewidth]{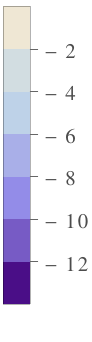}
\caption{\label{f:zz}%
(Color online) Density and contour plots of the connected correlator $\left\langle \sigma^+_o \sigma^-_\delta\right\rangle$. The data are for long-range XXZ chains with $\lvert\Lambda\rvert=40$ sites and exponents $\alpha=3/4$, $3/2$, and 3 (from left to right). Top: original data plotted in the $(\delta,t)$-plane, showing supersonic propagation for $\alpha=3/4$ and $3/2$, and a linear cone for $\alpha=3$. Middle row: logarithm of the original data plotted in the $(\ln\delta,\ln t)$-plane. Odd/even effects make it difficult to discern the shape of the causal region. Bottom row: as in the middle row, but with the odd/even effects removed by plotting the quantity defined in \eqref{e:max}.
}
\end{figure}

In addition to the connected correlation function $\left\langle \sigma^z_o \sigma^z_\delta\right\rangle_\text{c}$ discussed in the main text, we also computed the correlation function $\left\langle \sigma^+_o \sigma^-_\delta\right\rangle$ for the same parameter values. The original data are plotted in the top and middle rows of Fig.\ \ref{f:zz}, the bounded data (i.e., with odd/even effects removed) in the bottom row. Interestingly, the slope of the signals in the log-log plots is smaller than in $\left\langle \sigma^z_o \sigma^z_\delta\right\rangle_\text{c}$, and the possible leakage of information out of the cone for $\alpha=3$ appears to be more pronounced for $\left\langle \sigma^+_o \sigma^-_\delta\right\rangle$. Summarizing, these results further corroborate the conclusions drawn in the main text.

\bibliography{LRLR5}

\end{document}